\documentclass[10pt,conference]{IEEEtran}

\newif\ifanonymize
\ifanonymize \def\hl#1{}\else
\let\hl\relax
\fi

\IEEEoverridecommandlockouts
\usepackage{ascmac,amsmath}
\usepackage{graphicx}
\usepackage[utf8]{inputenc}
\usepackage{latexsym}
\usepackage{multirow}
\usepackage{url}
\usepackage{xcolor} 
\usepackage{amsfonts}

\usepackage{cite}
\usepackage{ascmac}
\usepackage{subfigmat}

\begin{document}
\title{An Experience Report on Regression-Free Repair of Deep Neural Network Model}
\hl{
\author{
\IEEEauthorblockN{Takao Nakagawa}
\IEEEauthorblockA{\textit{Fujitsu Limited}\\
Kawasaki, Japan \\
nakagawa-takao@fujitsu.com}
\and
\IEEEauthorblockN{Susumu Tokumoto}
\IEEEauthorblockA{\textit{Fujitsu Limited}\\
Kawasaki, Japan \\
tokumoto.susumu@fujitsu.com}
\and
\IEEEauthorblockN{Shogo Tokui}
\IEEEauthorblockA{\textit{Fujitsu Limited}\\
Kawasaki, Japan \\
tokui.shogo@fujitsu.com}
\and
\IEEEauthorblockN{Fuyuki Ishikawa}
\IEEEauthorblockA{\textit{National Institute of Informatics}\\
Tokyo, Japan \\
f-ishikawa@nii.ac.jp}
}}

\maketitle

\thispagestyle{plain}
\pagestyle{plain}

\begin{abstract}

Systems based on Deep Neural Networks (DNNs) are increasingly being used in industry.
In the process of system operation, DNNs need to be updated in order to improve their performance.
When updating DNNs, systems used in companies that require high reliability must have as few regressions as possible.
Since the update of DNNs has a data-driven nature, it is difficult to suppress regressions as expected by developers.
This paper identifies the requirements for DNN updating in industry and presents a case study using techniques to meet those requirements.
In the case study, we worked on satisfying the requirement to update models trained on car images collected in Fujitsu assuming security applications without regression for a specific class.
We were able to suppress regression by customizing the objective function based on NeuRecover, a DNN repair technique.
Moreover, we discuss some of the challenges identified in the case study.
\end{abstract}


\section{Introduction}
\label{sec:introduction}

The development of systems using deep neural networks (DNNs) differs from conventional system development because it is built in a data-driven manner.
DNN developers collect large amounts of data, label it with the correct answer, build the dataset, and train the model using the dataset.
The trained DNN models cannot achieve 100\% accuracy, i.e., not all tests pass.
If a failure is found during testing or operation that would result in a significant business loss, the developer or operator must fix the DNN model.
The fix usually involves retraining. Retraining requires data for correction and does not always correct the desired defects.
A DNN model repair technique is needed for this purpose.

There are three main types of DNN model repair techniques.
\textit{Training Program-centric Repair}
is an approach to detect and correct errors in the training program\cite{zhang2021autotrainer,wardat2022deepdiagnosis}.
\textit{Data-centric Repair}
is an approach to repair by training on data that has been augmented to match the repair target, or by training on data that has been selected as suitable for repair\cite{sensei2020gao,yu2021deeprepair}.
\textit{Model-centric Repair}
is an approach that repairs a trained model by changing the weights according to the target of the repair\cite{zhang2019apricot,sohn2022arachne}.

We developed a model-centric repair technique NeuRecover\cite{tokui2022neurecover},
which achieves conservative repair with less regression by narrowing the scope of repair to a small number of weights so as not to affect already observed successful cases.

However, existing studies have not reported that it is possible to repair faults without causing any regression at all, even though it is possible to reduce regression~\cite{srivastava2020empirical, yan2021positive}.
In this paper, we report a case study in which we built a DNN model from car image data collected by a company assuming security applications, and tried to repair it without regression using NeuRecoverLite, a simplified version of NeuRecover.
As a result of running 10 times each with 66 hyperparameters, we found a pattern that could be fixed without regression.
We also report the lessons learned from the case study.


\section{Regression-controlled Repair Technique: NeuRecoverLite}
\label{sec:neurecover}
A DNN repair technique, NeuRecover, has been proposed to repair DNN models locally without retraining by changing their parameters (weights) in search methods~\cite{tokui2022neurecover}.
NeuRecover detects data that regress from correct to wrong predictions (regression data) and data that improve from wrong to correct predictions (improvement data) during the training process of the DNN model, performs fault localization to identify parameters (weights) of the DNN model that do not affect the improvement data but only the regression data, and then performs a particle swarm optimization~\cite{pso1995James,pso2007Andreas} is used to search for values of the weights that reduce faults while preventing regression.
NeuRecoverLite is a simplified version of NeuRecover that performs localization without the results of the training process.
In the case study, we employed NeuRecoverLite because the training history was not being recorded in the field.

NeuRecoverLite consists of two stages: fault localization and particle swarm optimization.
Fault localization identifies weights that affected failed data but not passed data in order to find suspicious weights that should be repaired.
In particle swarm optimization, the weights identified in fault localization are optimized based on the fitness function in a direction that increases the number of data for which the misclassification is repaired while sustaining the number of passed data.

\subsubsection{Fault Localization}
\label{subsubsec:neurecover_localize}
NeuRecoverLite performs fault localization in two steps.
\begin{description}
    \setlength{\itemindent}{9pt}
    \setlength{\labelsep}{15pt}
    \item[Step\,i] Impact calculation
    \item[Step\,i\hspace{-1pt}i] Weight localization by set operation
\end{description}


In Step \,i, four degrees of impact of the weights to identify the weights to be repaired, the backward impact $\textit{back}_{\textrm{failed}}$ and forward impact $\textit{fwd}_{\textrm{failed}}$ on the failed data, and the backward impact $\textit{back}_{\textrm{passed}}$ and forward impact $\textit{fwd}_{\textrm{passed}}$ on the passed data, are computed.
The backward impact $\textit{back}$ is the weight $w_{i,j}$ connecting the $j$th neuron in the layer to be repaired to the $i$th neuron in the layer one step before, and the loss gradient of $w_{i,j}$ relative to the output $o_j$ in the layer one step before the target layer is $\frac{\partial L}{\partial w_{i,j}} = \frac{\partial L}{\partial o_j} \frac{\partial o_j}{\partial w_{i,j}}$.
The forward impact $\textit{fwd}$ is the value obtained by multiplying the output $o_i$ of the previous layer by the weight $w_{i,j}$, that is, the value $o_i \cdot w_{i,j}$ before nonlinear transformation by the activation function.

In Step\,i\hspace{-1pt}i, the weights are sorted by the four impacts to obtain a set of the top $N_g$ weights $B_{\textrm{failed}}$, $F_{\textrm{failed}}$, $B_{\textrm{passed}}$, $F_{\textrm{passed}}$ are retrieved.
The set of weights that affect the failed data and do not affect the passed data are the weights to be optimized according to $W_\textrm{localized} = (B_{\textrm{failed}} \cap F_{\textrm{failed}}) \backslash (B_{\textrm{passed}} \cap F_{\textrm{passed}})$.

\subsubsection{Particle Swarm Optimization}
\label{subsubsec:neurecover_optimize}
In a similar fashion to NeuRecover, NeuRecoverLite uses particle swarm optimization~\cite{pso1995James,pso2007Andreas} to optimize the weights localized by fault localization to repair misclassifications in DNN models.
Particle swarm optimization is known to be effective for optimization in continuous space and is suitable for repairing unbounded weights in the range of real numbers.

NeuRecoverLite uses the vector $\vec{x}$, which is the set of weights identified in the fault localization, to represent the position of the particles in the particle swarm optimization.
The fitness function of NeuRecoverLite increases as failed data is repaired, decreases as passed data is regressed, and increases as the value of each loss decreases.
As a difference from NeuRecover, NeuRecoverLite generates the initial value $\vec{x}_0$ of the particle vector from a normal distribution determined from the distribution of weights for half of the particles, and uses the original weight values for the other half of the particles, making it less likely to cause regression.

\section{Field Requirements of DNN Repair}
\label{sec:requirements}
Since DNN systems are developed in a data-driven manner, the methods, costs, and results obtained for repairing faults are also different from those of conventional systems.
However, many customers are not IT experts, and their requirements may be the same as those for conventional systems, without taking into account the characteristics of DNN systems. Therefore, the challenge for the field is to respond to quality requirements that could be achieved with conventional systems but cannot be easily achieved with DNN models.
One example of this issue is the suppression of regression after repair, which is tackled in this paper.

In the repair of DNN models, there are multiple levels of regression depending on the granularity, which is different from the conventional system.
\begin{enumerate}
    \item Overall accuracy level regression suppression
    \item Instance-level regression suppression
\end{enumerate}
The overall accuracy level regression suppression means that the overall accuracy of the data to be evaluated is not lower than before the repair.
Instance-level regression suppression means that the correct instances of the class to be repaired in the data to be evaluated do not change to failures after the repair.
These granularities are required at different levels by different users and products.

\section{Case Study}
\label{sec:experiment}
This section presents a case study on a proof-of-concept experiment on repairing with suppressed regression within Fujitsu using NeuRecoverLite.

\subsection{Research Questions}
\label{subsec:research_questions}
We set the following research question on repair without regression using NeuRecoverLite.

\begin{description}
    \item[RQ1:] Is it possible to repair without regression for a specific class of data?
    \item[RQ2:] Is it possible to repair without regression on data with drift?
    \item[RQ3:] How to set each hyperparameter to reduce regression?
\end{description}

RQ1 is to find out if it is possible to repair a particular class without regression in order to improve the performance of that class when it is not performing as expected.
RQ2 is a special case of RQ1 and examines whether a class can be repaired without regression to improve its performance when the data distribution changes during operation, for example, when a class that rarely appears at first becomes more frequent after a certain period of time.
In RQ3, through the experiments in RQ1 and RQ2, we will investigate the effect on the results when the settings of each hyperparameter are changed.

\subsection{Experiment Setup}
\label{subsec:setup}
In the case study, we built a neural network that classifies car images into seven classes.
The architecture is based on 27 convolutional layers and one fully connected layer.
The dataset consists of 10,186 images collected by Fujitsu, which are divided into training data, validation data, repair data, and test data, assuming updates during operation.
Input to the tool is passed samples of training data and failed samples of specific classes in the repaired data.

\begin{figure}[htbp]
    \begin{subfigmatrix}{2}
        \subfigure[Partially occluded]{\includegraphics[keepaspectratio, scale=0.5]{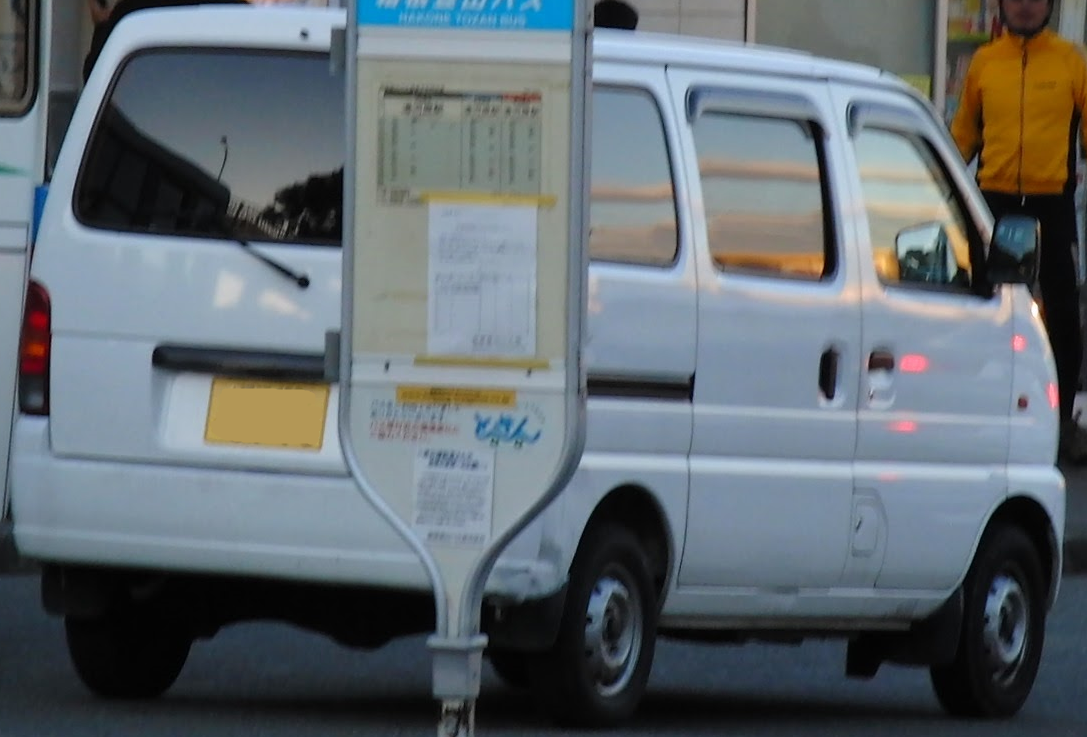}}
        \subfigure[Front/Back]{\includegraphics[keepaspectratio, scale=0.5]{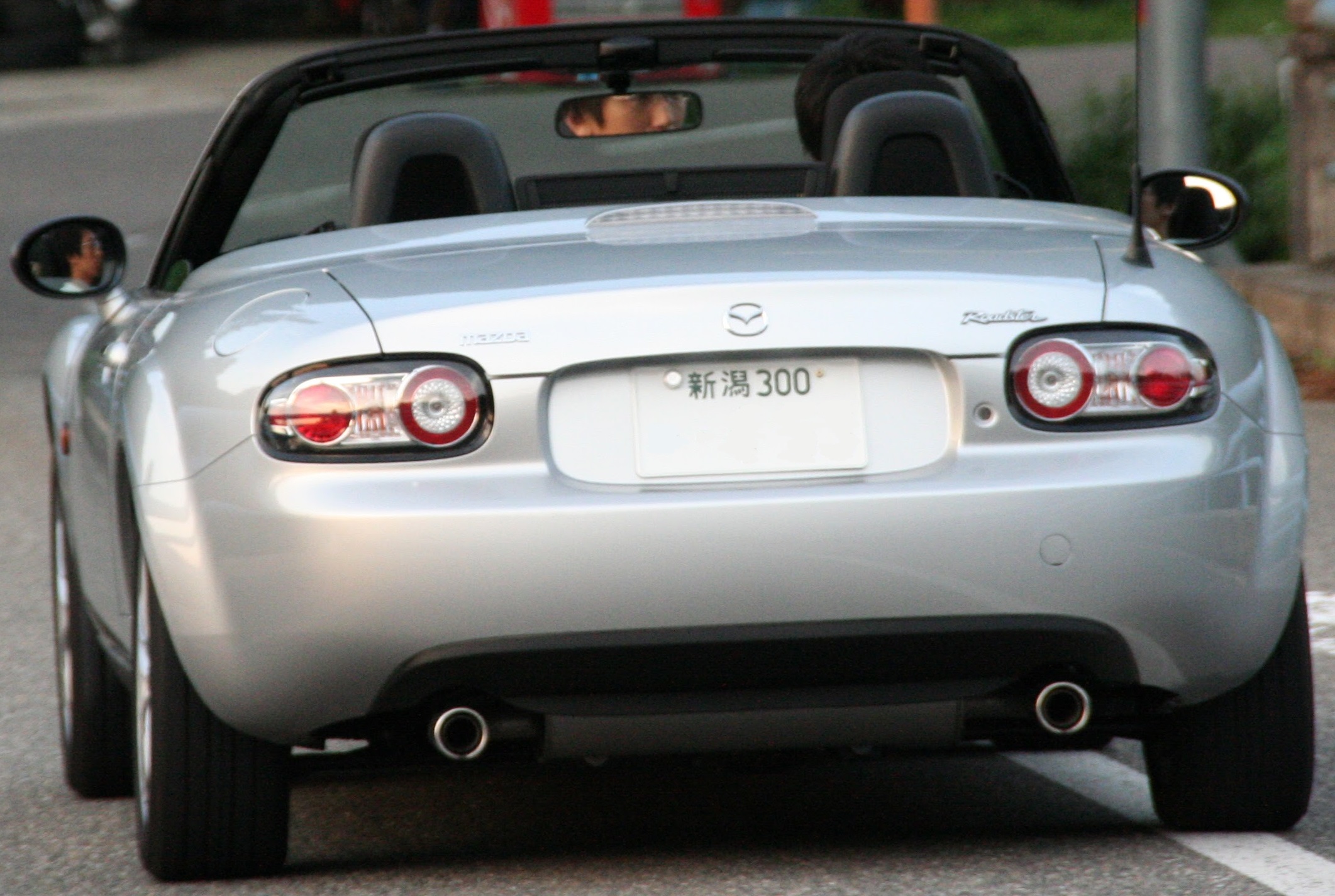}}
    \end{subfigmatrix}
    \caption{Hypothetical images of repair target class}
    \label{fig:targetclass}
\end{figure}

In this case study, we experimented with two cases of classes to be repaired: partial shielding and front/back (EXP-A, EXP-B). These classes are different from the classes to be classified.
Hypothetical examples of partial shielding and front/back images are shown in Figure \ref{fig:targetclass}.

We also prepared a case in which a pseudo-drift occurred and examined whether the drift could be repaired (EXP-C).
The procedure was designed to significantly reduce the percentage of classes to be repaired during training and to increase the percentage of classes to be repaired during the repair.

Table \ref{tab:experimentcond} summarizes the subjects of this experiment.

\begin{table}[htbp]
    \centering
    \caption{Experiment conditions}
    \label{tab:experimentcond}
    \begin{tabular}{c|c|c}
        ID    & Repair target class & Drift \\\hline
        EXP-A & Partially occluded & No  \\
        EXP-B & Front/Back         & No  \\
        EXP-C & Front/Back         & Yes
    \end{tabular}
\end{table}

The experiment was conducted while manually searching for hyperparameters to repair the DNN without regression.
Therefore, we did not exhaustively try all hyperparameter patterns, and as for the number of hyperparameter patterns, we tried 66 patterns in all.
Ten runs were performed for each hyperparameter setting, and the average of the results is examined.

The hyperparameters to be changed mainly are as follows.
\begin{enumerate}
    \item \textit{Fitness function}
    \item \textit{Intact weight ($\alpha$)}  
    \item \textit{Perfect intact (pi)}
    \item \textit{\# of localized weights (\#lw)}
    \item \textit{The number of positive samplings (\#pos)}
    \item \textit{The number of particles (\#p)}
\end{enumerate}

\textit{Fitness function} is the objective function used in particle swarm optimization, and it can be expressed as the following two formulas, (\ref{eq:fit1}) and (\ref{eq:fit2}).
Let $I_{neg}$ be the set of failed data to be repaired and $I_{pos}$ be the set of passed data sampled randomly at a fixed number.
$N_{patched}$ is the number of data in $I_{neg}$ that changed to passed data, and $N_{intact}$ is the number of data in $I_{pos}$ that did not change to failed data.
The $L$ is the loss function in the pre-repair model, and $L'$ is the loss function in the post-repair model.
The $\alpha$ is the \textit{Intact weight}, a hyperparameter to adjust the degree of suppression of regression.
The $\beta$ is a hyperparameter to adjust the effect of the loss function, which in this paper is set to $0.25$.
The $\delta$ is a small positive constant.
\begin{eqnarray}
    fitness &=& \frac{N_{\textrm{patched}}}{|I_{\textrm{neg}}|} + \alpha \cdot \frac{N_{\textrm{intact}}}{|I_{\textrm{pos}}|} \nonumber \\
            & & + \frac{L'(I_{neg}) + \delta}{L(I_{neg}) + \delta} + \frac{L'(I_{pos}) + \delta}{L(I_{pos}) + \delta} \label{eq:fit1} \\
    fitness &=& \frac{N_{\textrm{patched}}}{|I_{\textrm{neg}}|} + \alpha \cdot \frac{N_{\textrm{intact}}}{|I_{\textrm{pos}}|} + \beta  \cdot \frac{L'(I_{neg}) + \delta}{L(I_{neg}) + \delta} \label{eq:fit2}
\end{eqnarray}

\textit{pi} is a boolean value to set fitness to 0 if even one regression occurs.
The evaluated value $fitness'$ can then be expressed as in the following formula.
\begin{equation}
    fitness' =
    \begin{cases}
        0 & (pi = True \land N_{\textrm{intact}} / |I_{\textrm{pos}}| < 1) \\
        fitness & (otherwise)
    \end{cases} \nonumber
\end{equation}

Note that when \textit{pi} is true or $\alpha$ has a large value, these parameters contribute significantly to suppressing regression because they work in the direction of lowering the score when regression does occur.
The term on loss was introduced to evaluate cases where no maintenance/modification is seen. Equation (1) may neglect modifications when a loss reduction is easy for successful data, so Equation (2), which removes the loss term on successful data, was also prepared.

\textit{\#lw} is the number of localized weights that will be the output in localization, and how many of them to choose can be controlled when choosing the top $N_g$ weights in localization.

\textit{\#pos} is the number of instances sampled from the passed data to be used in the evaluation of the fitness function.
The larger the number, the closer it is to the entire passed data.

\textit{\#p} is the number of particles used in particle swarm optimization.
It is considered that the more there are, the more diverse the particles become during the search, and the larger the search space becomes.

\subsection{Results}
\label{subsec:results}
The hyperparameters of what was distinctive in the experiment are shown in Table \ref{tab:exp1results}.
\begin{table}[htbp]
    \centering
    \caption{Hyperparameters for each experiment}
    \label{tab:exp1results}
    \begin{tabular}{c|c|r|c|r|r|r}
        ID    & fitness         & $\alpha$ & pi & \#lw & \#pos & \#p \\ \hline
        EXP-A1 & (\ref{eq:fit1}) & 6       & False & 309  & 500  & 100 \\ 
        EXP-B1 & (\ref{eq:fit1}) & 6       & False & 282  & 500  & 200 \\ 
        EXP-C1 & (\ref{eq:fit2}) & 8       & True  & 32   & 500  & 200 \\ \hline 
        EXP-A2 & (\ref{eq:fit1}) & 6       & False & 309  & 500  & 200 \\ 
        EXP-A3 & (\ref{eq:fit1}) & 5       & False & 309  & 500  & 200 \\ 
        EXP-A4 & (\ref{eq:fit1}) & 4       & False & 309  & 500  & 200 \\ 
        EXP-B2 & (\ref{eq:fit1}) & 5       & False & 282  & 500  & 200 \\ 
        EXP-B3 & (\ref{eq:fit1}) & 4       & False & 282  & 500  & 200 \\ \hline 
        EXP-B4 & (\ref{eq:fit1}) & 6       & False & 282  & 1000 & 200 \\ \hline 
        EXP-C2 & (\ref{eq:fit2}) & 8       & True  & 76   & 500  & 200 \\ 
        EXP-C3 & (\ref{eq:fit2}) & 8       & True  & 53   & 500  & 200 \\ 
        EXP-C4 & (\ref{eq:fit2}) & 8       & True  & 16   & 500  & 200    
    \end{tabular}
\end{table}

\begin{figure}
    \centering
    \includegraphics[width=0.85\linewidth]{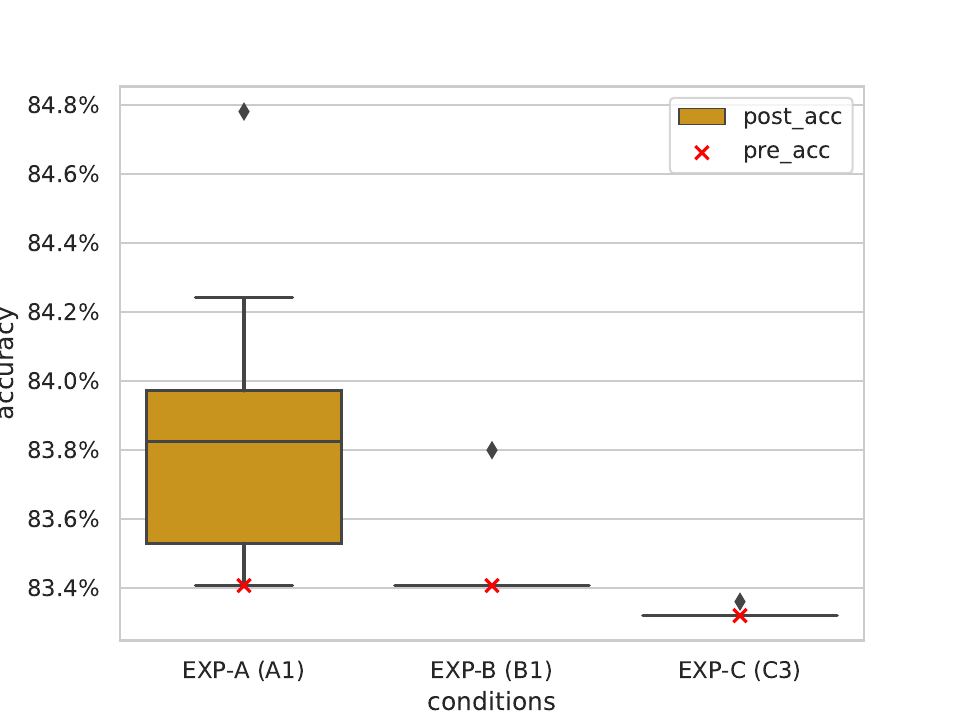}
    \caption{Accuracy of the minimum regression cases on test data}
    \label{fig:acc_minreg}
\end{figure}

\begin{figure}
    \centering
    \includegraphics[width=0.85\linewidth]{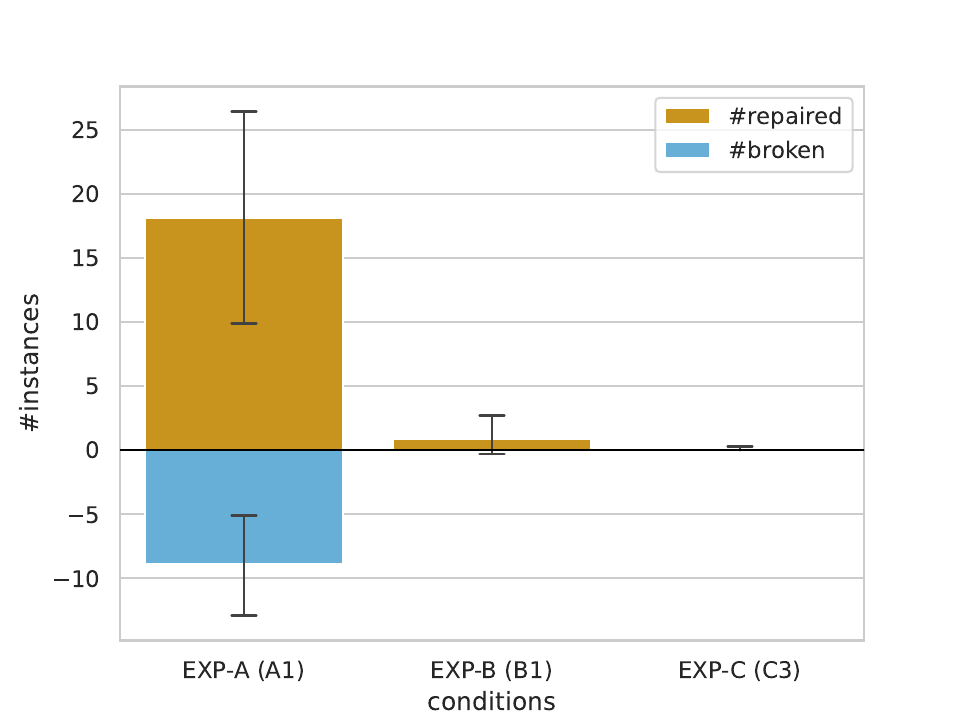}
    \caption{The number of broken/repaired instances of the minimum regression cases on test data}
    \label{fig:broken_repaired_minreg}
\end{figure}

Figure \ref{fig:acc_minreg} shows the accuracy of each experiment in minimizing the number of regressions, and Figure \ref{fig:broken_repaired_minreg} shows the number of regressions and repairs in each experiment.

In EXP-A, we achieved regression suppression at the overall accuracy level, but not at the instance level.
In EXP-B, we were able to suppress regression in both respects, although the improvement in overall accuracy was small.

\begin{itembox}[l]{Answer to RQ1}
It is possible to suppress regression at the overall accuracy level. On the other hand, whether regression can be suppressed at the instance level depends on the target. Even if regression can be suppressed, it is difficult to increase the number of repairs.
\end{itembox}

In EXP-C, we were able to suppress both regression at the overall accuracy level and the instance level. On the other hand, the improvement in overall accuracy was even less than in EXP-B.

\begin{itembox}[l]{Answer to RQ2}
It is possible to suppress regression at the overall accuracy level and at the instance level. On the other hand, it is more difficult to increase the number of repairs than in the absence of drift.
\end{itembox}

\begin{figure}
    \centering
    \includegraphics[width=0.9\linewidth]{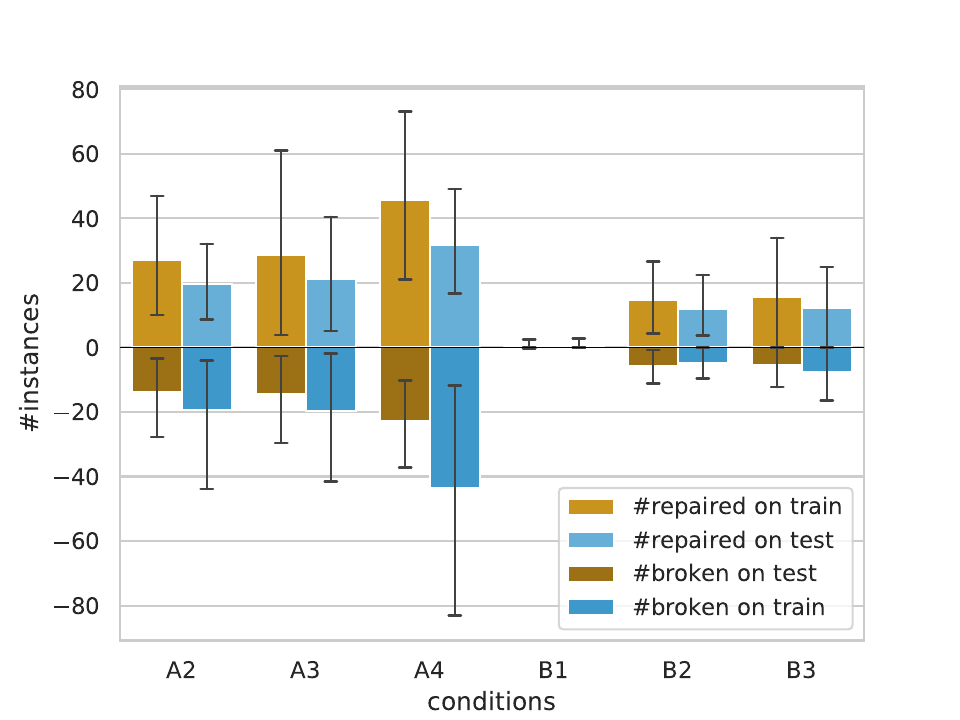}
    \caption{The number of broken/repaired instances for comparing EXP A/B and intact weights}
    \label{fig:broken_repaired_expab}
\end{figure}

Figure \ref{fig:broken_repaired_expab} shows the results of the number of regressions and repairs for EXP-A and EXP-B when only the intact weights ($\alpha$) were changeable and the other hyperparameters were left unchanged.
A2, A3, and A4 are for EXP-A when $\alpha$ is 6, 5, and 4, respectively.
Similarly, B1, B2, and B3 are for EXP-B when $\alpha$ is 6, 5, and 4, respectively.
In both EXP-A and EXP-B, the number of regressions can be reduced by increasing $\alpha$.
As the number of regressions decreases, the number of repairs also decreases accordingly.
Furthermore, even with the same hyperparameter settings, EXP-A and EXP-B show different regression and repair tendencies, with EXP-A showing a relatively higher number of regressions.

\begin{figure}
    \centering
    \includegraphics[width=0.9\linewidth]{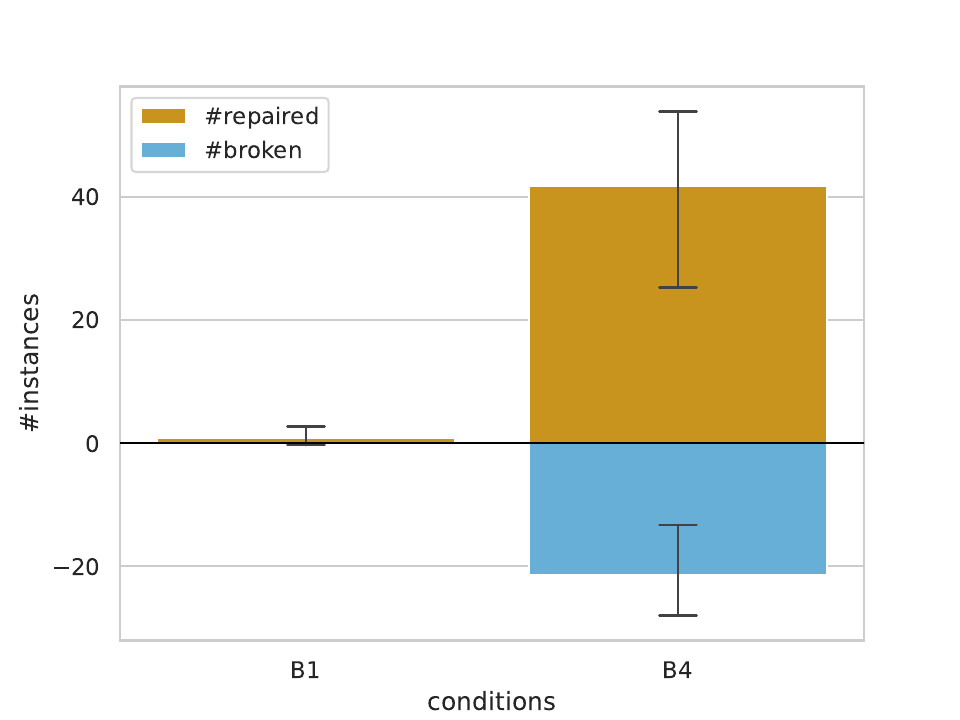}
    \caption{The number of broken/repaired instances for comparing \# of positive sampling}
    \label{fig:broken_repaired_npos}
\end{figure}

Figure \ref{fig:broken_repaired_npos} shows the number of regressions and repairs in EXP-B when the number of positive samplings is changed.
B4 was run with twice as many positive samplings (1000) as B1, but B4 had a higher number of regressions than B1.

\begin{figure}
    \centering
    \includegraphics[width=0.9\linewidth]{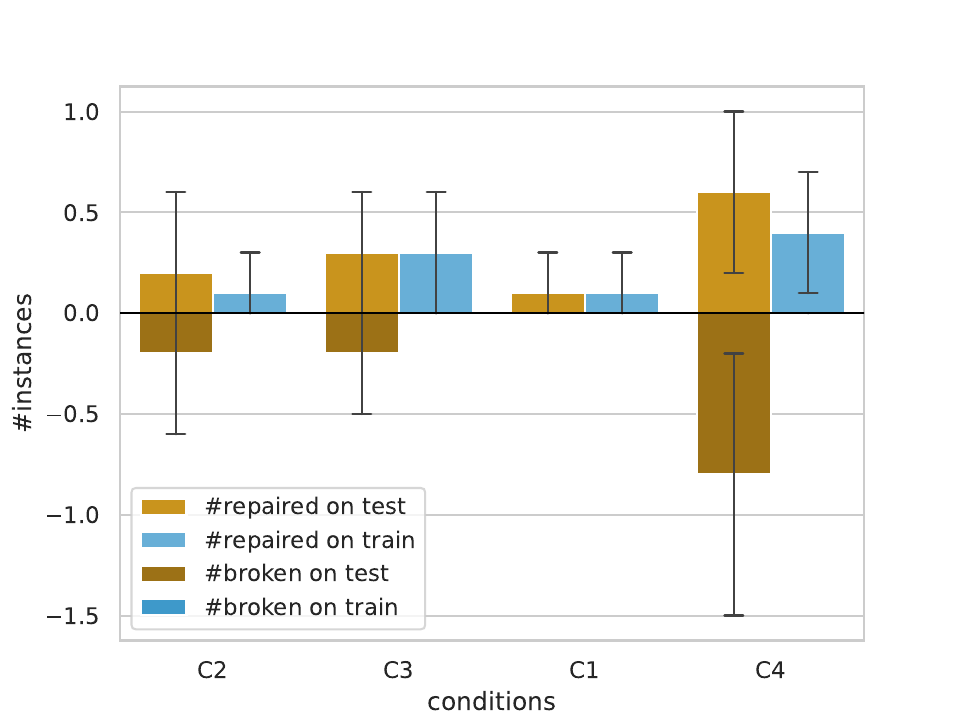}
    \caption{The number of broken/repaired instances for comparing \# of localized weights}
    \label{fig:broken_repaired_nlocalized}
\end{figure}

Figure \ref{fig:broken_repaired_nlocalized} shows the number of regressions and repairs when the number of localized weights is changed in EXP-C.
C2 has the highest number of localized weights and C4 has the lowest.
C1 has the lowest number of regressions, indicating that a certain amount of the number of localized weights is necessary, but further increase does not contribute to reducing the number of regressions.

\begin{itembox}[l]{Answer to RQ3}
The larger the intact weight($\alpha$), the more effective it is in suppressing the number of regressions.
If the number of positive samplings is too large, regression is more likely to occur.
The number of localized weights is more or less likely to cause regression.
\end{itembox}

\section{Lessons Learned}
\label{sec:lessons}


\subsection{Guaranteeing no regression on test data is difficult}
Repair can only be performed on known data, and suppressing regression on unknown data is inherently difficult.
Since the field may require assurance that at least known problems have been solved, there is still value in suppressing regression on training/validation data.

\subsection{Repairing large amounts of data without regression is difficult}
When attempting instance-level complete suppression of regression on both training and test data, the number of repairs will be very small (e.g., 0 or 1).
In some cases, hyperparameters resulted in weights that were almost unchanged from the original weights.
We believe that an optimization method that can completely suppress regression while ensuring search space would be useful.

\subsection{Difficulty of no regression depends on the target to be repaired}
EXP-A and EXP-B differ in the class to be repaired, while EXP-B and EXP-C differ in the presence of data drift. The results for each condition showed different trends, which may have been influenced by the difficulty of classifying the data to be repaired.
For example, the target of EXP-A (occluded objects) had more difficulty suppressing regression than EXP-B due to the existence of various patterns of occlusion in terms of size, location, etc.
Also, there were fewer successful repairs in EXP-C than in EXP-B, suggesting that data drift has increased the difficulty of repairs.

In the future, it may be necessary to consider appropriate repair techniques depending on the target to be repaired.

\subsection{Repair difficulty leads to unintended bias in results}
In the fitness function, the value per repair is considered equal for all cases, regardless of the difficulty or importance of each case to be repaired.
Inevitably, priority is given to cases that are easier to repair or have more similar cases, resulting in biased repair results.
However, in field application, there may be cases where priority is given to specific data, and cases that are rare or difficult to repair may also be considered.
For the future, we need repair techniques that takes into account the difficulty and importance of each case to be repaired.


\subsection{How to select hyperparameters}
Since the intact weight is a value that expresses how much importance is placed on the number of regressions, it is natural that the larger the weight, the more the number of regressions can be suppressed.
On the other hand, the greater the number of positive samplings, the greater the number of data used, which intuitively seems to contribute to reducing regression, but in this experiment, it worked in the opposite direction, increasing regression.
We believe that this is due to the fact that when the number of positive samplings is large, the second term of the fitness function, $\alpha \cdot N_{intact}/|I_{pos}|$, indicates that the effect of fitness per regression is smaller.
In other words, since the choice of hyperparameters may change significantly depending on the design of fitness, it is important how to design a better fitness function as a prerequisite for good hyperparameter selection.

\section{Conclusion}
\label{sec:summary}
This paper demonstrated how to completely suppress regression through a case study of repair for car image classification.

Future work includes increasing the amount of repair while completely suppressing regression.
In addition, proof of effectiveness in tasks other than the image classification task will help to find further applicability.
Furthermore, reducing the number of hyperparameters to be adjusted will lead to greater applicability in the field.

\hl{
\section*{Acknowledgment}
    This work was partly supported by JST-Mirai Program Grant Number JPMJMI20B8, Japan.
}

\bibliographystyle{IEEEtran}
\bibliography{references}
\end{document}